\documentclass[aps,prb,twocolumn,showpacs]{revtex4}
\usepackage{amsmath,graphicx}

\begin{document}

\title{Optical properties of 1D photonic crystals based on
multiple-quantum-well structures}

\author{M. V. Erementchouk}
\author{L. I. Deych}
\author{A. A. Lisyansky}

\affiliation{Physics Department, Queens College, City University
of New York, Flushing, New York 11367, USA}

\begin{abstract}
A general approach to the analysis of optical properties of
photonic crystals based on multiple-quantum-well structures is
developed. The effect of the polarization state and a
non-perpendicular incidence of the electromagnetic wave is taken
into account by introduction of an effective excitonic
susceptibility and an effective optical width of the quantum
wells. This approach is applied to consideration of optical
properties of structures with a pre-engineered break of the
translational symmetry. It is shown, in particular, that a layer
with different exciton frequency placed at the middle of an MQW
structure leads to appearance of a resonance suppression of the
reflection.
\end{abstract}


\maketitle

\section{Introduction}

Structures with spatially modulated dielectric properties
(photonic crystals) attract an ever growing interest since the
first papers where they were considered.\cite{Yablonovitch,John}
This interest is caused by unique opportunities that such
structures provide, to affect, in a controllable way, fundamental
\emph{microscopic} processes of light-matter interaction through a
modification of \emph{macroscopic} geometric characteristics of
the structures. This makes such structures of obvious interest not
only for fundamental physics but also for applications. Most of
the works devoted to photonic crystals considered structures made
of materials with a frequency independent dielectric
constant.\cite{photonic_crystals,Yariv_book} Recently, however, a
new class of structures, which can be described as resonant or
optically active photonic crystals, has attracted particular
attention.\cite{Kuzmiak_Maradudin,DeychLivdan,Nojima:2000,%
excitons_grating,Raikh_braggariton,Loncar_laser,Tikhodeev_PRL,%
Huang_Joannopoulos_PRB:2003,Huang_Joannopoulos_PRL:2003}
In these structures periodic modulation of the dielectric constant
is accompanied by the presence of internal excitations of
constituent materials resonantly interacting with light within a
certain frequency region, and resulting in a strong frequency
dispersion of constituent dielectric constants. Extreme cases of
such structures are so called optical lattices, in which well
localized resonant elements are periodically distributed through
the medium with a uniform dielectric constant. Originally, the
concept of optical lattices referred to structures formed by cold
atoms,\cite{DeutschOpticalLattices} but it was also applied to a
special kind of multiple quantum well structure (MQW), which were
considered as a semiconductor analog of a one-dimensional optical
lattice.\cite{Ivchenko_Pikus}

MQW is a periodic multilayer structure built of two semiconductor
materials, for instance, GaAs and Al${}_x$Ga${}_{1-x}$As, in which
electrons and holes are confined in narrower layers of a material
with a smaller band gap (quantum wells) separated by wide layers
of a semiconductor with larger band-gap (barriers). In this case,
the role of dipole active resonant excitations is played  by
excitons confined to respective quantum wells, and if the width of
the barriers is large enough, excitons from different quantum
wells do not interact directly. They, however, still can interact
through their common radiation field, and in this sense they are
similar to atomic optical lattices. This analogy is, however,
exact only if one can neglect a difference in refractive indexes
of wells and barriers. This approximation was widely used in most
papers devoted to long-period MQW structures, in which the period
of the structure is comparable with the wavelength of exciton
radiation.\cite{IvchenkoMQW,Ivchenko1991} Of special interest are
so called Bragg structures, in which the excitonic wavelength is
in Bragg resonance with the periodicity, and which are
characterized by a significantly enhanced radiative coupling
between quantum well excitons. As a result of this coupling, light
propagates through such a structure in the form
exciton-polaritons, whose dispersion law is characterized by two
branches with  a band-gap between
them\cite{Ivchenko1991,IvchenkoMQW}. The width of this stop band
is significantly enhanced compared to off-Bragg structures, and
this is what makes such structures of particular interest for
applications.

In realistic MQW structures, however, dielectric constants of the
wells and barriers are not equal to each other, and the presence
of resonant optical excitations is accompanied by a periodic
modulation of the background dielectric constant. These
structures, therefore, represent a special case of one-dimensional
resonant photonic crystals, optical properties of which are
characterized by an interplay between interface reflections and
resonant light-exciton interaction. The effects of the refractive
index contrast on the optical properties of MQW structures have
not been, of course, overlooked in previous studies. In
particular, a modification of the Bragg condition and reflection
spectra at normal incidence of Bragg MQWs in the presence of the
contrast have been discussed in Refs.~%
\onlinecite{IvchenkoContrast,JointContrast}. The effects of the
dielectric mismatch on optical properties of single quantum wells
\cite{Kavokin_oblique,AndreaniPRB1992} or an MQW structure
embedded in a dielectric environment\cite{DielEnv} was also taken
into account. However, a complete analytical theory of MQW based
photonic crystals has not yet been developed. While optical
spectra of any given MQW based structure can be easily obtained
numerically, this is not sufficient when one needs to design a
structure with predetermined optical properties, which is a key
element in utilizing these structures for optoelectronic
applications.  The main difficulty of this task is the presence of
a large number of experimental parameters such as an angle of
incidence, a polarization state, indices of refraction, widths of
the barriers and the quantum wells, etc, which are in a
complicated way related to spectral characteristics of a
structure. In order to resolve this difficulty, one needs a
general effective analytical approach that would facilitate
establishing relationships between material parameters and
spectral properties of MQW based structures for an arbitrary angle
of incidence and polarization state of incoming light. In the
present paper we develop such a method and apply it to a case of
MQW structure with an intentionally broken periodicity (an MQW
structure with a ``defect"). The method is based on a transfer
matrix approach and consists of two steps. In the first step we
show that a quantum well embedded in a dielectric environment can
be described in exactly the same way as a quantum well in vacuum
by introducing an effective excitonic susceptibility and an
effective optical width of the quantum well layer. In the second
step we establish relations between these effective quantum well
characteristics and parameters of a total transfer matrix,
describing propagation of light throughout an entire structure.
The method is rather general and can be applied to a great variety
of different MQW structures with light of an arbitrary
polarization, incident at an arbitrary angle. In order to
demonstrate the power of our approach, we consider reflection
spectra of an MQW structure in which a central well is replaced
with a well having a different resonant frequency. Such structures
have been considered previously in a number of papers
\cite{DefectAPL,DefectMQW,OmegaDefectPRB} in the optical lattice
approximation. Here we show that the presence of the refractive
index contrast does not destroy the remarkable reflection
properties of such structures, confirming, therefore, their
potential for optoelectronic applications.

\section{A single quantum well in a dielectric environment}

Propagation of the electromagnetic wave in structures under
discussion is governed by the Maxwell equation
\begin{equation}\label{eq:Maxwell_equations}
  \nabla \times \nabla \mathbf{E} = \frac{\omega^2}{c^2}[
  \epsilon_\infty(z) \mathbf{E} + 4\pi \mathbf{P}_{exc}],
\end{equation}
with modulated background dielectric permeability,
$\epsilon_\infty(z)$, which is assumed to take values $n_b^2$ and
$n_w^2$ in the barriers and the quantum wells materials
respectively. For the sake of concreteness we assume hereafter
that $n_w > n_b$, unless otherwise explicitly specified. $P_{exc}$
is the excitonic contribution to the polarization and is defined
by
\begin{equation}\label{eq:exciton_polarization}
  \mathbf{P}_{exc} = \chi(\omega) \int \Phi(z) \Phi(z') \mathbf{E}(z')d z',
\end{equation}
where $\Phi(z)$ is the exciton envelope function. Here we have
restricted ourselves by taking into account $1s$ heavy-hole
excitons only and have neglected the in-plane dispersion of the
excitons. The frequency dependence of the excitonic susceptibility
is described by
\begin{equation}\label{eq:susceptibility_ideal}
  \chi(\omega) = \frac{\alpha}{\omega_0 - \omega - i\gamma},
\end{equation}
where $\omega_0$ is the exciton resonance frequency, $\gamma$ is
the non-radiative decay rate of the exciton, $\alpha = \epsilon_b
\omega_{LT}a_B^3\omega_0^2/4c^2$, $\omega_{LT}$ is the exciton
longitudinal-transverse splitting and $a_B$ is the bulk exciton
Bohr's radius.

Due to the absence of an overlap of the exciton wave functions
localized in different quantum wells and the linearity of the
Maxwell equations, the propagation of the electromagnetic wave
along the structure can be effectively described by a
transfer-matrix. Using the usual Maxwell boundary conditions the
transfer matrix through one period of the structure in the basis
of incoming and outgoing plane waves can be written in the form
\begin{equation}\label{eq:transfer_matrix_period}
  T = T_b^{1/2} T_{bw} T_w T_{wb} T_b^{1/2},
\end{equation}
where
\begin{equation}\label{eq:transfer_matrix_barrier}
  T_b^{1/2} =
  \begin{pmatrix}
    e^{i \phi_b/2} & 0 \\
    0 & e^{-i \phi_b/2}
  \end{pmatrix}
\end{equation}
is the transfer matrix through the halves of the barriers
surrounding the quantum well. Here $\phi_b = \omega n_b d_b
\cos\theta_b/c$ with $d_b$ being the width of the barrier and
$\theta_b$ being an angle between the wave vector $\mathbf{k}$
inside the barrier and the direction of the $z$-axis,
$\hat{\mathbf{e}}_z$.

\begin{figure}
  \includegraphics[width=2.9in, angle=0]{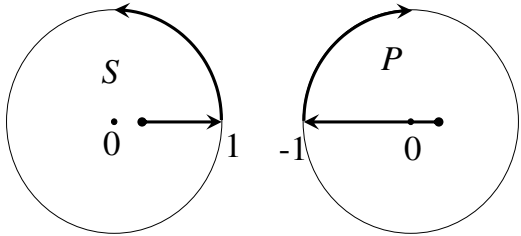}\\
  \caption{A scheme of the angular dependence of the Fresnel
  coefficients $\rho_s$ and $\rho_p$ upon the angle of incidence,
  $\theta_b$, on the complex plane is shown. At normal
  incidence the coefficients have values shown by small filled circles
  (the same for both polarizations). When the angle of incidence
  increases the coefficients follows the arrows on the lines.
  $\rho_p$ passes through $0$ at the Brewster's angle, both
  coefficients reach the unit circle at the angle of total
  internal reflection.
}\label{fig:rho_ps}
\end{figure}

The scattering of the electromagnetic wave at the interface
between the quantum well and the barrier caused by the mismatch of
the indices of refraction of their materials is described by
\begin{equation}\label{eq:transfer_matrix_barrier_well}
  T_{bw} = T_{wb}^{-1} =T_\rho(\rho)\equiv \frac{1}{1 + \rho}
  \begin{pmatrix}
    1 & \rho \\
    \rho & 1
  \end{pmatrix},
\end{equation}
where $\rho$ is the Fresnel reflection coefficient. The interface
scattering depends upon both the angle of incidence of the wave
and its polarization state. These effects are effectively
described by Fresnel coefficients (see, e.g. Ref.
\onlinecite{BornWolf}) $\rho_s$ and $\rho_p$
\begin{eqnarray}\label{eq:Fresnel_coefficients_vals}
 \rho_s = \frac{n_w \cos\theta_w - n_b \cos\theta_b}{n_w \cos\theta_w + n_b
 \cos\theta_b},
 \nonumber \\
 \rho_p = \frac{n_w \cos\theta_b - n_b \cos\theta_w}{n_w \cos\theta_b + n_b
 \cos\theta_w}
\end{eqnarray}
for $s$ ($\mathbf{E} \perp (\mathbf{k},\hat{\mathbf{e}}_z)$) and
$p$ ($\mathbf{E} \parallel (\mathbf{k},\hat{\mathbf{e}}_z)$)
polarizations respectively. The angular dependence of these
coefficients upon the angle of incidence measured inside the
barrier is schematically shown in Fig. \ref{fig:rho_ps}.

Finally,
\begin{equation}\label{eq:transfer_matrix_well}
  T_w = \begin{pmatrix}
    e^{i\phi_w}(1-i S) & -i S \\
    i S & e^{-i\phi_w}(1 + i S)
  \end{pmatrix}
\end{equation}
is the transfer matrix through the quantum well. Here $\phi_w =
\omega n_w d_w\cos \theta_w/c$, where $d_w$ is the width of the
quantum well and $\theta_w$ is the angle between $\mathbf{k}$
inside the quantum well and $\hat{\mathbf{e}}_z$. The excitonic
contribution to the scattering of the light is described by the
function
\begin{equation}\label{eq:S_def}
  S = \frac{\Gamma_0}{\omega - \omega_0 + i\gamma},
\end{equation}
which we will call the excitonic susceptibility in what follows.
The radiative decay rate, $\Gamma_0$, at normal incidence is
determined by
\begin{equation}\label{eq:Gamma_def}
  \Gamma_0 = \frac{1}{2}\pi k\omega_{LT}a_B^3
   \left(\int \Phi(z) \cos kz\, dz\right)^2.
\end{equation}

For oblique incidence the radiative decay rates are renormalized
in different ways for different polarizations. For
$p$-polarization in addition to this renormalization  it is also
necessary to take into account a possible splitting of $Z$- and
$L$-exciton modes \cite{Andreani_SSC,Citrin_Lifetime,Citrin_MQW}
that gives rise to a two-pole form of $S$. However, in materials
with the zinc-blend structure, the $Z$-mode of the heavy-hole
excitons is optically inactive, and one can describe angle
dependencies of the radiative decay rate for $s$- and
$p$-polarizations respectively by simple expressions
\begin{equation}\label{eq:Gamma_sp_oblique}
  \Gamma_0^{(s)} = \Gamma_0 /\cos \theta_w,
  \qquad \Gamma_0^{(p)} = \Gamma_0 \cos \theta_w.
\end{equation}

Thus, propagation of light in the structures under consideration
depends upon a number of natural parameters such as Fresnel
coefficients, exciton frequencies and radiative decay rate, and
optical widths, which (with the exception of $\omega_0$) depend
upon the angle of incidence of the wave and its polarization
state.

Our next step will be to simplify the presentation of the total
transfer matrix through the period of the structure in such a way
that makes the relations between the elements of the transfer
matrix and the natural parameters of the structure more apparent.
The most complicated part of the transfer matrix is the product
$T_{bw}T_wT_{wb}$, which describes the reflection of the wave from
the interface and its interaction with quantum well excitons. We
simplify it by noting that this product can be presented as
$T_{bw}T_wT_{wb}=\widetilde{T}_w$, where $\widetilde{T}_w$ has the
same form as $T_w$, Eq.~(\ref{eq:transfer_matrix_well}), but with
renormalized parameters
\begin{equation}\label{eq:transfer_matrix_effective_well}
  \widetilde{T}_w=T_{bw}T_wT_{wb} = \begin{pmatrix}
    e^{i\tilde\phi_w}(1-i \tilde S) & -i \tilde S \\
    i \tilde S & e^{-i\tilde\phi_w}(1 + i \tilde S)
  \end{pmatrix},
\end{equation}
where the effective excitonic susceptibility, $\tilde S$, and the
phase shift, $\tilde\phi_w$, are defined as
\begin{eqnarray}\label{eq:effective_susceptibility_and_width}
  \tilde S  = S \frac{1+ \rho^2 - 2 \rho \cos \phi_w}{1- \rho^2}
   + 2\rho\frac{\sin\phi_w}{1-\rho^2}, \nonumber
  \\
  e^{i(\tilde\phi_w - \phi_w)} = \frac{1 - \rho e^{-i\phi_w}}{1 - \rho
  e^{i\phi_w}}.
\end{eqnarray}
Here $\rho$ denotes the Fresnel coefficient for the wave of a
respective polarization. Taking into account the diagonal form of
the transfer matrix through the barrier $T_b$ one can see that the
total transfer matrix through the period of the structure again
has the form of a single quantum well transfer matrix and is
determined by (\ref{eq:transfer_matrix_well}) where the phase
$\tilde\phi_w$ is replaced by a total phase $\phi = \phi_b +
\tilde\phi_w$.

Thus, we have shown that the propagation of the wave in MQW
based photonic crystals can be described in terms of properties of
a respective optical lattice with renormalized parameters. The
renormalization of the phase is the simplest one: the expression
for $\tilde\phi_w$ can be rewritten as
\begin{equation}\label{eq:effective_phase}
  \tilde\phi_w = \phi_w \frac{1+\rho}{1-\rho}
\end{equation}
provided that the change of phase $\phi_w$ across the well is much
smaller than $2\pi$, which is usually true for long period MQW
structures. Hence, one of the effects of the index of refraction
contrast is reduced to a simple renormalization of the optical
width of the quantum well.

The effective susceptibility, $\tilde{S}$, consists of two terms.
One of them has a singularity at the exciton frequency while the
second remains smooth in a wide frequency region. The relation
between these terms depends upon the frequency and near the
exciton resonance the second term is negligibly small. The
frequency region where the nonsingular addition to the effective
susceptibility can be neglected is determined by
\begin{equation}\label{eq:condition_small_regular_addition}
  |\omega - \omega_0| < \omega_{\mathrm{min}} =
  \frac{\Delta_\Gamma^2}{2\Delta_{PC}}\frac{1-\rho}{1+\rho} ,
\end{equation}
where $\Delta_\Gamma = \sqrt{2\Gamma_0\omega_0/\pi}$ is the
half-width of the forbidden gap in a Bragg MQW without a mismatch
of the indices of refraction, and $\Delta_{PC} = 2\omega_r \rho
\sin[\phi_w(\omega_0)]/\pi(1-\rho^2)$ is the half-width of the
forbidden gap in a passive photonic crystal characterized by the
same value of the mismatch calculated with the assumption of the
narrowness of the gap in comparison with  the position of the
center of the gap, $\omega_r$.

It follows from Eq.~(\ref{eq:condition_small_regular_addition})
that the resonant part of the effective susceptibility,
Eq.~(\ref{eq:effective_susceptibility_and_width}), makes the main
contribution  over the entire frequency region of exciton
polaritons stop band when
\begin{equation}\label{eq:excitons_beat_photons}
  \Delta_\Gamma > 2 \Delta_{PC}.
\end{equation}

\begin{figure}
  \includegraphics[width=3.5in, angle=0]{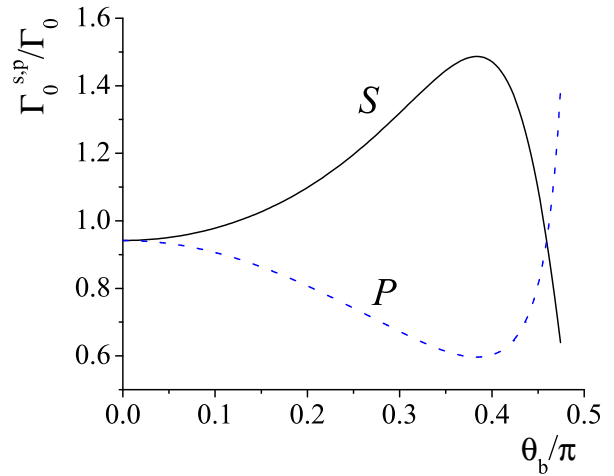}\\
  \caption{Change of the effective radiative decay rates $\Gamma_{s}$ (solid line)
  and $\Gamma_p$ (dotted line) with
  the angle of incidence for a single quantum well.}\label{fig:effective_Gamma_angle}
\end{figure}

In realistic multiple GaAs/Al${}_x$Ga${}_{1-x}$As structures the
values $\rho < 0.03$ of the Fresnel coefficient at normal
incidence\cite{Adachi} and $\phi_w(\omega_0) \sim 0.1 \pi$ can be
considered as typical, so both quantities $\Delta_\Gamma$ and
$\Delta_{PC}$ are of the same order of magnitude $\sim 10^{-2}$
eV. Therefore, in principle, both signs of the inequality in
Eq.~(\ref{eq:excitons_beat_photons}) are possible. Here, however,
we consider the simplest case when one can neglect the
non-singular term in $\tilde S$ and approximate the latter by
\begin{equation}\label{eq:effective_parameters_small}
 \tilde S = S \frac{1-\rho}{1+\rho}.
\end{equation}
In this approximation, the effect of the contrast is reduced to an
additional modification of the radiative decay rate (or,
consequently, the oscillator strength). Depending upon the value
of the Fresnel coefficient one can observe either an enhancement
(when $\rho<0$) or a reduction (when $\rho>0$) of exciton
radiative recombination. Since usually $n_w> n_b$, the Fresnel
coefficient for the normal incidence is positive and therefore the
oscillator strength is diminished compared to the case of the
absence of the contrast. When the angle of incidence increases, in
addition to different dependencies of the Fresnel coefficients
corresponding to different polarization states following from
Eqs.~(\ref{eq:Fresnel_coefficients_vals}), it is necessary to take
into account direct modification of the oscillator strength given
by Eqs.~(\ref{eq:Gamma_sp_oblique}).
Fig.~\ref{fig:effective_Gamma_angle} shows the dependence of the
factor modifying the radiative decay rate upon the angle of
incidence.

\section{Optical properties of MQW structures}

The reduced description presented in the previous section works
well within a frequency region satisfying the inequality given in
Eq.~(\ref{eq:condition_small_regular_addition}). For a single
quantum well (or even for short MQW structures) and for angles not
too close to the angle of total internal reflection, the condition
(\ref{eq:condition_small_regular_addition}) is fulfilled for the
entire frequency region of interest, which, in this case, is of
the order of magnitude of $\Gamma_0$. Therefore, under these
circumstances the mismatch of the indices of refraction can be
treated perturbatively, so that all results neglecting the
mismatch remain valid up to renormalization of the oscillator
strength. However, when the number of quantum wells in an MQW
structure increases, the frequency region affected by excitons
widens almost linearly,\cite{IvchenkoMQW} and, therefore, the
regular contribution to the effective susceptibility becomes
important, and more detailed analysis is necessary.

Keeping in mind subsequent application to more complicated
structures, it is convenient to introduce a special representation
for a transfer matrix
\begin{equation}\label{eq:transfer_matrix_bq}
  T(\theta, \beta) = \begin{pmatrix}
    \cos \theta - i\sin\theta \cosh\beta & - i\sin\theta \sinh\beta \\
    i\sin\theta \sinh\beta & \cos \theta + i\sin\theta \cosh\beta
  \end{pmatrix},
\end{equation}
where the parameters of the representation, $\theta$ and $\beta$,
are related to the ``material" parameters $S$ and $\phi_w$
entering the transfer matrix by
\begin{eqnarray}\label{eq:bq_parameters}
  \cos \theta = \mathrm{Tr}\, T/2 = \cos\phi + S\sin\phi,
  \nonumber \\
  \coth\beta = \cos\phi - S^{-1}\sin\phi.
\end{eqnarray}
This representation is valid for an arbitrary system that
possesses a mirror symmetry with respect to a plane passing
through the middle of the structure. It can be easily derived
taking into account the equality of the determinant of the matrix
to one, and the circumstance that the mirror symmetry requires
off-diagonal elements to be imaginary. Due to the general
character of the representations (\ref{eq:transfer_matrix_bq}),
the material parameters entering Eq.~(\ref{eq:bq_parameters}) can
be either the parameters of a single quantum well,
Eq.~(\ref{eq:transfer_matrix_well}) or the effective parameters
$\tilde S$ and $\tilde \phi$,
Eqs.~(\ref{eq:transfer_matrix_effective_well}) and
(\ref{eq:effective_susceptibility_and_width}), of a barrier-well
sandwich, or even parameters characterizing the entire MQW
structure as long as the latter possess the mirror symmetry.

Using this representation we can introduce the following
transformation rule for transfer matrices
\begin{equation}\label{eq:bq_algebra}
    T_H(\psi) T(\theta,\beta) T_H^{-1}(\psi)= T(\theta,\beta +
  2\psi),
\end{equation}
where matrix $T_H$ describes a hyperbolic rotation with a dilation
and has the form of
\begin{equation}\label{eq:T_H_psi_definition}
  T_H(\psi) =
 e^{\psi} \begin{pmatrix}
   \cosh\psi & -\sinh\psi \\
   -\sinh\psi & \cosh\psi
 \end{pmatrix}.
\end{equation}
This transformation rule can be used, for instance, for
diagonalization of transfer matrices, which can be achieved by
choosing parameter $\psi=-\beta/2$. Matrix $T_H$ can be turned
into matrix $T_{bw}$, Eq.~(\ref{eq:transfer_matrix_barrier_well}),
which describes propagation of waves through interface between two
media with different refraction coefficients by using the
following relation between $\psi$ and the Fresnel parameter
$\rho$: $\rho = -\tanh(\psi)$ (a detailed discussion of a relation
between interface scattering and the hyperbolic rotation can be
found in Ref.~\onlinecite{Monzon_Fresnel_rotation}). This means
that the transformation, Eq.~(\ref{eq:bq_algebra}), can be either
used to describe the interface between two different structures,
or in order to present any type of non-diagonality of the
transfer-matrix as resulting from some effective interface. With
the help of Eq.~(\ref{eq:bq_algebra}), any symmetric multilayer
structure can be replaced by a uniform slab with the width given
by $\theta$ and the index of refraction determined by $\psi$.
Therefore, it can be used to describe structures which are more
complicated than a simple three layer barrier-well sandwich
considered in the previous section. For instance, using
Eq.~(\ref{eq:bq_algebra}) we can immediately derive an expression
for the transfer matrix $T_N$ of a sequence of identical blocks
described by $T(\theta,\beta)$:
\begin{equation}\label{eq:bq_algebra_N_power}
T_N = T(\theta,\beta)^N = T(N\theta,\beta).
\end{equation}
Because the reflection from the structure described by the
transfer matrix $T$ given in the basis of incoming and outgoing
waves is
\begin{equation}
r=-T_{21}/T_{22},
\end{equation}
for a structure containing $N$ blocks we have
\begin{equation}\label{eq:reflectivity_M_general}
  r_N = -\frac{i \sinh \beta}{\cot N\theta + i \cosh \beta}.
\end{equation}
Eq. (\ref{eq:reflectivity_M_general}) in the case of $\Gamma_0 =
0$ reproduces the result well-known for a passive multilayer
structure.\cite{Yariv_book,Yariv_paper,BENEDICKSON:1996}

To find a relation between the quantities entering this expression
and the elements of the transfer matrix through the period of the
structure it is enough to multiply both the numerator and the
denominator by $\sin\theta$ and to use
Eq.~(\ref{eq:bq_parameters}). If each block is characterized by an
effective susceptibility $\tilde S$ and the phase $\phi=\phi_b
+\tilde\phi_w$ then we obtain for the reflection coefficient
\begin{equation}\label{eq:reflectivity_M_general_step2}
  r_N = \frac{i \tilde S}{\cot(N\lambda)\sin \lambda  + i(\tilde S \cos\phi - \sin\phi)}.
\end{equation}

\begin{figure}
  \includegraphics[width=3.5in, angle=0]{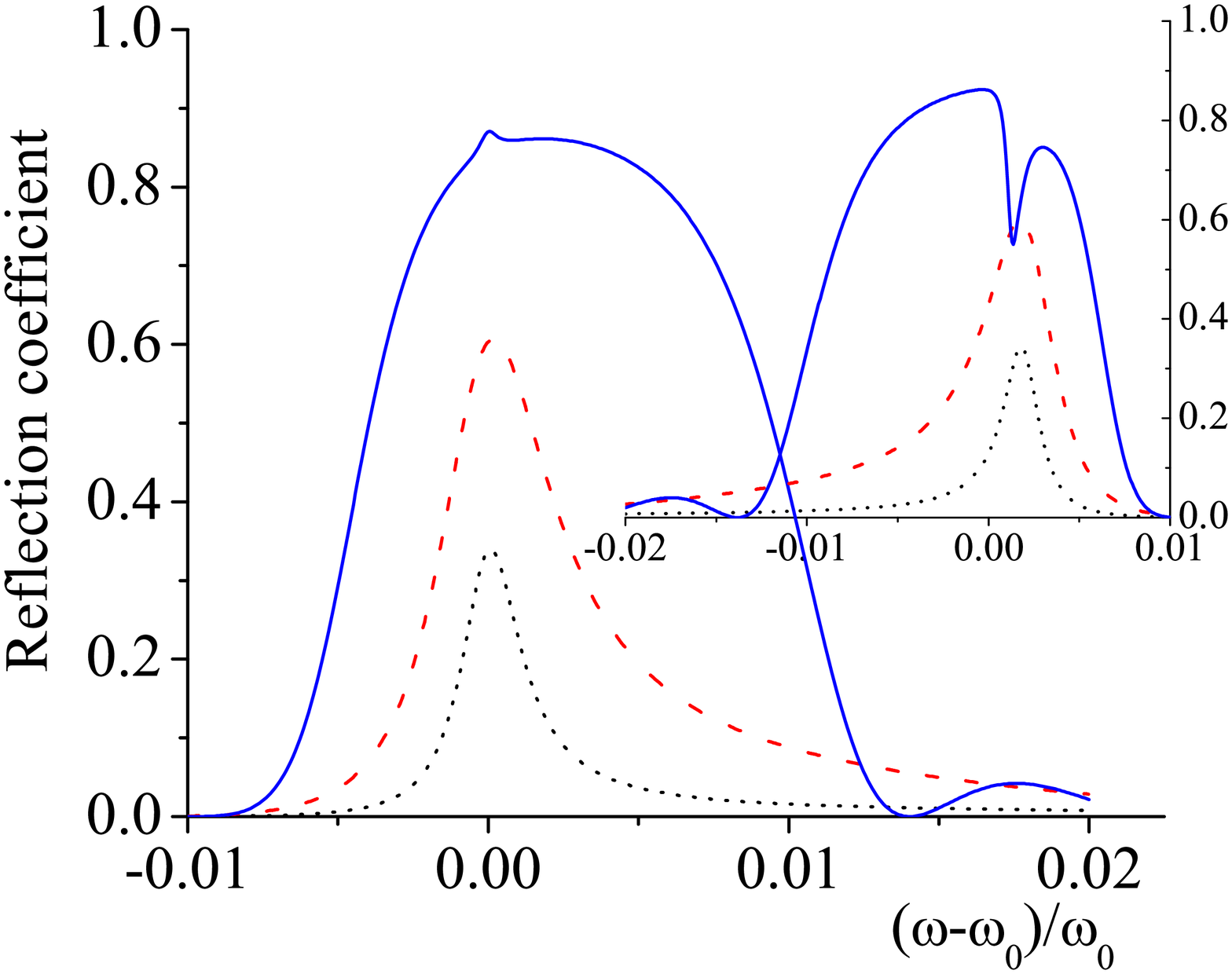}\\
  \caption{Dependence of the amplitude reflection coefficient $|r_N|^2$ upon the
  frequency. The main plot shows the reflection of structure
  satisfying
  the modified Bragg condition with $\rho=0.005$, the homogeneous broadening $\gamma = 500$ $\mu$eV
  and the length $N=10,25,100$ (dot, dash, solid lines, respectively).
  On the inset reflection spectra of structures that satisfy
  the standard Bragg condition are shown. To make the
  correspondence with the results of Ref.~\onlinecite{KhitrovaMQW}
  clearer we chose $\rho = -0.005$.
 }\label{fig:mqw_reflection}
\end{figure}

We analyze the reflection coefficient for not too long structures,
such that $N \ll N_c = \lambda^{-1}$, which appear naturally in
most applications of MQW structures. In this approximation,
$\cot(N\lambda)\sin \lambda \approx N$ and the reflection can be
written directly in terms of the material parameters
\begin{equation}\label{eq:reflectivity_M_linear}
  r_N = \frac{i N \tilde S}{1 + iN(\tilde S \cos\phi - \sin\phi)}.
\end{equation}
The amplitude of reflection, $|r|^2$, has the typical form shown
in Fig.~\ref{fig:mqw_reflection}. It is characterized by a strong
reflection band around the exciton frequency, which is a
manifestation of the strong resonant exciton-light interaction. It
is interesting to note that $|r_N|^2$ goes to $0$ at
$\omega=\omega_0 - \omega_{\mathrm{min}}$ [see
Eq.~(\ref{eq:condition_small_regular_addition})] that is at the
frequency where $\tilde S = 0$. The reflection has an asymmetric
form since it is a sum of two terms: one of them is even with
respect to the frequency $\omega_0$ and the second is odd. The
latter is due to nonzero mismatch and in the approximation used
above does not vanish at infinity. Both these terms have a typical
width
\begin{equation}\label{eq:exciton_peculiarity_width}
  \delta = \frac{N\Gamma_0(1-\rho)^2 \cos\phi_+}{
  (1-\rho^2)^2 + N^2(\sin\phi_+ - \rho^2 \sin\phi_-)^2},
\end{equation}
where $\phi_\pm = \phi_b \pm \phi_w$. It should be noted that the
actual optical width of the quantum well determined by $\phi_w$
enters the definition of $\phi_\pm$, rather than the modified
$\tilde\phi_w$. We are interested in maximizing exciton related
effects in the reflection spectra of our structures, which means
designing structures with as large a width $\delta$ as possible.
One can see from Eq.~(\ref{eq:exciton_peculiarity_width}) that the
width demonstrates essentially non-monotonous dependence upon the
number of quantum wells in the structure: it grows linearly for
small $N$, but starts decreasing as $N^2$ for larger $N$. If the
coefficient in front of the $N^2$-term in the denominator of
Eq.~(\ref{eq:exciton_peculiarity_width}) vanishes, then the linear
growth of $\delta$ would go unchecked as long as $N$ does not
exceed $N_c$.  Thus, the condition for maximizing the excitonic
effects in the reflection spectrum can be formulated as
\begin{equation}\label{eq:reflection_max_linear}
\rho^2 = \sin\phi_+(\omega_0)/\sin\phi_-(\omega_0).
\end{equation}
If one neglects the mismatch of the indices of refraction, this
equation takes a well known form of a condition for the Bragg
resonance between the period of the structure and the exciton
radiation.\cite{Ivchenko1991} In the case of small $\rho$,
Eq.~(\ref{eq:reflection_max_linear})  can be approximated as
\begin{equation}\label{eq:Bragg_general}
\rho = \cos(\phi_+/2)/\cos(\phi_-/2),
\end{equation}
which coincides with a modified Bragg condition introduced in
Ref.~\onlinecite{JointContrast}, which actually requires that the
exciton frequency is equal to the low frequency boundary of the
passive (without excitons) photonic crystals' stop band.
Therefore, Eq.(\ref{eq:Bragg_general}) can be considered as the
most general condition defining Bragg MQW structures.

For Bragg MQWs the expression for the reflection coefficient,
Eq.~(\ref{eq:reflectivity_M_general_step2}), is essentially
simplified and can be approximated as
\begin{equation}\label{eq:reflectivity_M_Bragg}
r_N = \frac{i N \tilde S}{1 + i N \tilde S }.
\end{equation}
This expression gives a generalization of a well known result
about the linear dependence of the width of the exciton-polariton
reflection band on the number of quantum wells in Bragg MQW
structures.\cite{IvchenkoMQW}

When the system is detuned from the Bragg resonance, the
transparency window appears in the band gap. It shows up in the
form of a dip near the exciton frequency in the reflection
spectrum (see inset in Fig. \ref{fig:mqw_reflection}). For
example, in Ref.~\onlinecite{KhitrovaMQW} the reflection was
measured for MQW structures that satisfied the Bragg condition for
structures without contrast. In other words, the period of those
structures was made to coincide with the half-wavelength at the
exciton frequency calculated without taking into account photonic
crystal modification of light dispersion. These effects can
significantly modify the wavelength of light resulting in a
detuning of the structures studied in
Ref.~\onlinecite{KhitrovaMQW} from actual Bragg resonance. As a
result, spectra observed in that paper demonstrate features
specific  for slightly off-resonance structures [see
Fig.~(\ref{fig:mqw_reflection})].

\section{MQW structures with defects}

The results obtained in the previous sections are quite general
and can be applied directly to more complicated situations. As an
example, in this section we consider a reflection spectrum of a
system in which one of the barrier-well-barrier elements has
properties different from those of all other elements of the
structure. These structures can be described as MQW structures
with a ``defect." The break of the translational symmetry leads to
appearance of states localized near the defect layer with
frequencies in the forbidden gap of a host structure. Due to
resonance tunnelling of light via these local states, a
significant modification of the reflection spectrum occurs. This
effect was studied in regular passive one-dimensional photonic
crystals,\cite{Impurity_modes:1993,Liu:1997,Figotin:1998,Felbacq:2000,tunable_defect}
and in Bragg MQWs in the optical lattice
approximation.\cite{DeychPLA,DefectMQW} It was found, in
particular, that the defect affects the spectrum most strongly
when it is placed in the middle of the structure.\cite{DefectMQW}
In this case, the system demonstrates the mirror symmetry and the
results of the previous sections can be used. Indeed, for a
structure $ABA$ built of blocks $A$ and $B$ described by the
transfer matrices (\ref{eq:transfer_matrix_bq}) with parameters
$\theta_{A,B}$ and $\beta_{A,B}$, one has
\begin{equation}\label{eq:sandwich_transfer}
    T(\theta_A,\beta_A) T(\theta_B,\beta_B) T(\theta_A,\beta_A) =
    T(\theta,\beta).
\end{equation}
That is, the whole structure can also be described by the matrix
(\ref{eq:transfer_matrix_bq}) with
\begin{equation}\label{eq:theta_sandwich}
\begin{split}
\cos \theta = \cos \varphi_+\,\cosh^2 \delta\beta
  - \cos\varphi_-\,\sinh^2 \delta\beta, \\
  \coth(\beta-\beta_A) = \cos(2 \theta_1)\coth\delta\beta
  + 
  \frac{\sin \varphi_-}{\sin\theta_2 \,\sinh(2\delta\beta)},
\end{split}
\end{equation}
where $\varphi_\pm = 2\theta_1 \pm \theta_2$ and
$\delta\beta=(\beta_2 - \beta_1)/2$. Applying now relations
(\ref{eq:bq_parameters}) one can express the result in terms of
parameters $S$ and $\phi$, and use the results for reflection
described above.

In some particular cases, however, the problem of scattering of
light can be solved without resorting to the transformation rule
(\ref{eq:theta_sandwich}). Let us consider a situation when the
block $B$ is a single quantum well surrounded by barriers so it
can be described by the matrix similar to
(\ref{eq:transfer_matrix_effective_well}) with parameters $S_d$
and $\phi_d$. Let the block $A$ be an MQW structure described by
$\theta$ and $\beta$. Thus, the transfer matrix through the whole
structure is
\begin{eqnarray}\label{eq:simple_sandwich_transfer}
 T = T(\theta,\beta)T_\rho(\rho)
T(S_d,\phi)T_\rho^{-1}(\rho) T(\theta,\beta),
\end{eqnarray}
where $T_\rho(\rho)$ takes into account a possible mismatch of the
indices of refraction of the defect layer and the host and, $\rho$
is the corresponding Fresnel coefficient.

The transfer matrix (\ref{eq:simple_sandwich_transfer}) can be
simplified in several steps. First, as has been described before,
we can treat $\beta$ as an addition to the mismatch noting that
\begin{equation}
T_H^{-1}(\beta/2)T_\rho(\rho) = T_\rho(\tilde\rho),
\end{equation}
where $\tilde \rho =(\rho + \rho')/(1+\rho\rho')$ and
$\rho'=\tanh(\beta/2)$. Then, similar to
Eq.~(\ref{eq:transfer_matrix_effective_well}) we can introduce
effective quantities $\tilde S$ and $\tilde\phi$
\begin{eqnarray}\label{eq:simple_sandwich_S_and_phi}
   \tilde S  = S_d \frac{1+ \tilde\rho^2 - 2 \tilde\rho \cos \phi}{1- \tilde\rho^2}
    + 2\tilde\rho\frac{\sin\phi}{1-\tilde\rho^2}, \nonumber
  \\
  e^{i(\tilde\phi - \phi)} = \frac{1 - \tilde\rho e^{-i\phi}}{1 - \tilde\rho
  e^{i\phi}}.
\end{eqnarray}
The next step is a multiplication of $T(\tilde S,\tilde\phi)$ by
the diagonal matrices $T(\theta,0)$ what leads to a simple shift
of the phase, $T(\tilde S,\tilde\phi + 2\theta)$. Finally, the
terminating matrices, $T_H(\beta/2)$ and $T_H^{-1}(\beta/2)$, are
taken into account by modifying $\tilde S$ and $\tilde\phi$. Thus
the resultant transfer matrix $T$ takes the form
(\ref{eq:transfer_matrix_effective_well}), i.e. $T=T(\tilde{\tilde
S}, \tilde{\tilde{\phi}})$ with
\begin{eqnarray}\label{eq:simple_sandwich_effective_effective}
  \tilde{\tilde{S}}= \tilde S \frac{1+ \rho'^2 +
      2 \rho' \cos (\tilde\phi+2\theta)}{1- \rho'^2}
    - 2\rho'\frac{\sin(\tilde\phi +2\theta)}{1-\rho'^2}, \nonumber
  \\
  e^{i(\tilde{\tilde{\phi}} - \tilde{\phi}-2\theta)} =
   \frac{1 + \rho' e^{-i\tilde{\phi}-2i\theta}}{1 + \rho'
  e^{i\tilde\phi+2i\theta}}.
\end{eqnarray}
These expressions together with
Eq.~(\ref{eq:reflectivity_M_general}) give a complete solution of
the problem of propagation of light through the MQW structure with
an arbitrary defect in the middle.

One can consider several particular types of defects. An example
is a well with the exciton frequency different from the
frequencies of all other wells, an $\Omega$-defect. Another
possible example could be a defect element with the width of the
barriers different from the rest of the structure. It is
interesting to note that a standard optical microcavity with a
quantum well at its center can also be considered within the same
formalism. For example, after substitution of $\tilde{S}$ from
(\ref{eq:simple_sandwich_S_and_phi}) into Eq.
(\ref{eq:simple_sandwich_effective_effective}) one obtains an
expression that contains a singular term (proportional to $S_d$)
and regular terms. Choosing such widths of the barriers
surrounding the quantum well so that the regular terms vanish in
the vicinity of the exciton frequency one has the reflection
determined by the exciton susceptibility with renormalized
oscillator strength. The excitonic contribution to the scattering
in such a structure will not be obscured by the interface
scattering.

We demonstrate the application of the results obtained above by a
detailed consideration of an $\Omega$-defect. This type of defect
was analyzed in
Refs.~\onlinecite{DefectMQW,OmegaDefectPRB,DefectAPL} in the
scalar model for the electromagnetic wave in MQW structures
without a mismatch of the indices of refraction. It has been shown
there that in the presence of homogeneous and inhomogeneous
broadening of excitons, the effect of the defect is  prominent
when the frequency of the exciton resonance in the defect layer,
$\omega_d$, is close to the boundary of the forbidden gap in the
host system, and the length of the system is not too big. The
reflection spectrum in this case has the characteristic Fano-like
dependence with a minimum followed by a closely located maximum.
Such a spectrum makes this structure a potential candidate for
novel types of devices such as optical switches or
modulators.\cite{OmegaDefectPRB,DefectAPL} It is interesting,
therefore, to find out how the refractive index mismatch  affects
spectral properties of such a structure.

For the frequencies within the polariton stop-band of the host
structure one has: $\theta = M(\pi + i\lambda)$, where $M$ is the
number of quantum wells in the parts of the structures surrounding
the defect, and $M\lambda \ll 1$. The description becomes much
simpler if we assume that the width of the defect layer is tuned
to the Bragg resonance at the frequency $\omega_d$, that is if
$\phi(\omega_d) = \pi$. That makes the second term in the
expression for $\tilde S$,
Eq.~(\ref{eq:simple_sandwich_S_and_phi}), negligible in a wide
region of frequencies and the expression for $\tilde{\tilde{S}}$
takes a very simple form
\begin{equation}\label{eq:Omega_defect_SS}
  \tilde{\tilde{S}} = 2 M S_h + S_d\frac{1+\rho}{1-\rho},
\end{equation}
where $S_h$ is the effective excitonic susceptibility of the host,
given by an expression similar to
Eq.~(\ref{eq:effective_susceptibility_and_width}) with the exciton
frequency $\omega_0 = \omega_h$. In derivation of Eq.
(\ref{eq:Omega_defect_SS}) we have neglected the small term
$\propto S_d S_h$. The reflection coefficient can be obtained by
substituting this expression into Eq.
(\ref{eq:reflectivity_M_general}) with $N=1$. The reflection has
peculiarities near the exciton frequencies of the host and the
defect, and in the absence of broadening becomes $0$ at the
frequency where $\tilde{\tilde{S}} = 0$. Assuming that the
photonic contribution to the forbidden gap in the host is small,
$\Delta_{PC} < \Delta_\Gamma$, we can approximately find this
frequency as
\begin{equation}\label{eq:reflection_drop_general}
  \omega_R = \omega_d - \frac{\Delta_\omega}{2M + 1} -
  16\frac{M^2 \Delta_\omega^2 \Delta_{PC}}{\Delta_\Gamma^2
  (2M+1)^3},
\end{equation}
where $\Delta_\omega = \omega_d - \omega_h$ is the difference
between the defect and the host exciton frequencies, and we assume
for concreteness that $\omega_d> \omega_h$.

Setting the mismatch of the indices of refraction in this
expression to zero we reproduce the expression for $\omega_R$
obtained in Ref.~\onlinecite{OmegaDefectPRB}. The fact that the
contrast does not preclude the reflection coefficient from going
to zero at a certain point is not at all obvious because it might
have been expected that the interface reflection would set a limit
on the decrease of the reflection. However, as seen from
Eq.~(\ref{eq:reflection_drop_general}), the mismatch leads only to
an additional shift of the zero point away from the defect exciton
frequency. We would also like to comment on the dependence of
the resonance frequency on the angle of incidence and the
polarization of the electromagnetic wave. These characteristics
enter Eq.~(\ref{eq:reflection_drop_general}) through the Fresnel
coefficients, Eq.~(\ref{eq:Fresnel_coefficients_vals}), that
determine the photonic band gap, $\Delta_{PC}\propto
\rho/(1-\rho^2)$. Therefore angular and polarization dependencies
of the zero point of reflection follow the behavior of
$\Delta_{PC}$. The obtained expression for the reflection
coefficient also allows for analyzing the position of the maximum
of reflection, and its maximum magnitude, but the resulting
expressions turn out to be too cumbersome and we do not present
them here.

The formalism presented in this paper allows one to take into
account effects of homogeneous and inhomogeneous broadenings
because the main results obtained  do not use a particular form of
the excitonic susceptibility. For example, the reflection
coefficient of a structure with an inhomogeneous broadening can be
obtained in the effective medium approximation by using
Eq.~(\ref{eq:reflectivity_M_general}) with $\tilde{\tilde{S}}$
instead of $S$ and with inhomogeneously broadened $S_{h,d}$ in
Eq.~(\ref{eq:reflection_drop_general}):
\begin{equation}\label{ed:inhomogenous_broadening}
  S_{h,d} =\int d\omega_0 f_{h,d}(\omega_0)
   \frac{\Gamma_0}{\omega - \omega_0 +
  i\gamma}.
\end{equation}
where $f_{h,d}$ are the distribution functions of the exciton
frequencies in the host and defect layers.
respectively\cite{KavokinGeneral,KavokinSpectrum,OmegaDefectPRB}
However, as has been discussed in
Ref.~\onlinecite{OmegaDefectPRB}, if $\omega_R$ is far enough from
$\omega_d$, i.e. if $|\omega_R - \omega_d| \gg \Delta$, where
$\Delta$ is the inhomogeneous broadening, the effect of the latter
is negligible and the magnitude of the reflection is determined by
\begin{equation}\label{eq:magnitude_S_min}
  |\tilde{\tilde{S}}| \approx
  \frac{\pi \gamma(2M+1)}{4M \omega_d \Delta_\omega^2}
  \left[\Delta_\Gamma^2 (2M+1)^2 + 16M (2M-1)\Delta_{PC}\Delta_\omega
  \right]
\end{equation}
and is small provided the smallness of the homogeneous broadening,
$\gamma\ll\omega_d$, and $\Delta_\omega \gtrsim \Delta_\Gamma
\gtrsim \Delta_{PC}$.

A typical form of the reflection in the vicinity of this resonant
drop has been discussed in Refs.
\onlinecite{OmegaDefectPRB,DefectAPL}. Here we would like to note
that for multiple defect structures composed of several blocks $A$
considered above the reflection can be qualitatively described by
Eq. (\ref{eq:reflectivity_M_Bragg}) where $N$ is understood as the
number of such blocks. This expression shows that increasing $N$
can be interpreted as increasing an effective radiative decay
rate. This qualitatively explains the results of numerical
calculations in \onlinecite{DefectAPL} where it was found that the
maximal value of the reflection increases with $N$ while the value
of the contrast remains about the same.

\section{Conclusion}

In the present paper, the general problem of light propagation in
an MQW based photonic crystal characterized by both spatial
modulation of the dielectric constant, and dipole active exciton
states in quantum wells, is considered. It is shown that the
mismatch of indices of refraction between barriers and wells can
be taken into account by introduction of an effective excitonic
susceptibility and an effective optical width of the quantum
wells. The effective susceptibility has two terms, one of which is
almost independent of frequency, while the second is resonant in
nature. For short enough MQW structures (or in the vicinity of the
exciton frequency) the regular term can be neglected and the
effect of the mismatch of the indices of refraction reduces to a
modification of the excitonic oscillator strength. In a general
case, the reflection spectrum becomes essentially asymmetric and
non-trivially dependent upon the number of quantum wells in the
structure. It is shown that in order to obtain the strongest
exciton induced reflection band the structure must satisfy a
certain resonance condition. This is a Bragg resonance condition
between the period of the MQW structures and the wavelength of the
electromagnetic wave. The latter has to be calculated from a
dispersion law for a structure with the spatially modulated
refraction index.

The developed approach is applied to analysis of the reflection
spectrum of a structure with an intentionally introduced defect
element, which breaks the translational symmetry of a system. More
detailed analysis is carried out for a special kind of defect,
characterized by the presence of a layer in the middle of the
structure with a different frequency of the exciton resonance. It
is shown that the main characteristics of the reflection spectrum
of such structures obtained in the absence of the refraction index
contrast survive in the presence of the additional interface
reflections. In particular, a significant decrease  of the
reflection takes place even in the presence of the contrast, which
is important for possible applications of such structures.

\end{document}